\documentclass[a4,10pt]{article}
\usepackage{amsmath,amssymb, bm, statex, braket}
\usepackage{enumerate, fancybox}

\usepackage[dvipdfmx]{graphicx}

\allowdisplaybreaks[4]
\newcommand{\extr}[1]{\underset{#1}{\mathrm{extr}\,}}

\newcommand{\nn}{\nonumber \\}
\newcommand{\mcal}[1]{ \mathcal{#1} }
\newcommand{\mrm}[1]{ \mathrm{#1} }

\newcommand{\ave}[1]{\langle #1 \rangle}
\newcommand{\Ave}[1]{\Big\langle #1 \Big\rangle}

\newcommand{\trace}[1]{\mathrm{Tr}\,#1}

\begin{document}

\begin{center}
{\huge{\textsc{Perturbative Interpretation of Adaptive\\ \vspace{2mm}Thouless-Anderson-Palmer Free Energy}}}\\

\ \\
\ \\
{\Large{Muneki Yasuda\footnote{Corresponding author: muneki@yz.yamagata-u.ac.jp}$^*$, Chako Takahashi$^*$, and Kazuyuki Tanaka$^{\dagger}$}}\\
\ \\
$^*$Graduate School of Science and Engineering, Yamagata University, Yonezawa 992-8510, Japan\\
$^{\dagger}$Graduate School of Information Sciences Tohoku University, Sendai 980-8579, Japan
\end{center}

\subsubsection*{abstract:}
In conventional well-known derivation methods for the adaptive Thouless-Anderson-Palmer (TAP) free energy, special assumptions 
that are difficult to mathematically justify except in some mean-field models, must be made.
Here, we present a new adaptive TAP free energy derivation method. 
Using this derivation technique, without any special assumptions, 
the adaptive TAP free energy can be simply obtained as a high-temperature expansion of the Gibbs free energy.

\section{Introduction}\label{secIntro}

The adaptive Thouless-Anderson-Palmer (TAP) equation, which is obtained via the minimum condition of the adaptive TAP free energy (ATAPFE), 
can be used to solve some spin glass models, such as the Sherrington--Kirkpatrick model and 
the Hopfield model~\cite{Opper&Winther2001a,Opper&Winther2001b}. 
A message-passing-type algorithm for solving the adaptive TAP equation has been proposed in Ref.\cite{Yasuda2013}.
Two methods for deriving the ATAPFE are known.   
The first approach (i) is based on the cavity method, the linear response relation, and the Plefka expansion. 
In derivation (i), a crucial assumption must be made, i.e., the cavity distributions are taken to be Gaussian distributions with variances independent of the external fields~\cite{Opper&Winther2001a,Opper&Winther2001b}. 
The second approach (ii) is based on the Plefka expansion. 
In derivation (ii), a different crucial assumption is made. i.e., that the intractable terms in the expansion can be replaced by tractable terms originating from a Gaussian model~\cite{Opper&Winther2001b, AdaTAP_NIPS14}.

In this paper, we present a new method for deriving the adaptive TAP free energy, which is based on the Plefka expansion and the linear response relation. 
Our method appears similar to the derivations (i) and (ii) in many respects.
However, in the method proposed in this study, the ATAPFE is obtained via a high-temperature expansion of a Hessian matrix appearing in the true Gibbs free energy.  
Further, the ATAPFE can be obtained without any unnatural assumptions. (Note that a conventional assumption for a Hessian matrix is made). 
The proposed method can be expected to gain our understanding for the ATAPFE.

\section{Gibbs Free Energy of Ising Model}

On an undirected graph $G(V,E)$, the Ising model is defined in the form of the Gibbs distribution, $P(\bm{x}) \propto\exp[-\beta E(\bm{x})] / Z$, of the energy function
\begin{align}
E(\bm{x}) := -\sum_{i \in V} h_ix_i - \sum_{\{i,j\} \in E}J_{ij}x_ix_j
= - \bm{h}^{\mrm{T}}\bm{x}  - \bm{x}^{\mrm{T}}\bm{J} \bm{x}/ 2,
\label{eqn:energy}
\end{align} 
where $\bm{x} = \{x_i \in \{+1,-1\} \mid i \in V\}$ are the Ising variables.  
The $\bm{J} = \{J_{ij}\}$ are the symmetric interactions and there are no self-interactions, $J_{ij} = J_{ji}$ and $J_{ii} = 0$, 
and the $\bm{h} = \{h_i\}$ are the external fields. 
$Z_{\beta}$ and $\beta$ are the partition function and the (positive) inverse temperature, respectively. 
For a test distribution $Q(\bm{x})$, minimizing the variational free energy  
\begin{align*}
\mcal{F}[Q] := \beta \sum_{\bm{x}}E(\bm{x})Q(\bm{x}) + \sum_{\bm{x}}Q(\bm{x}) \ln Q(\bm{x})
\end{align*}
under the constraint, $m_i = \sum_{\bm{x}} x_i Q(\bm{x})$ $\forall i \in V$, yields the Gibbs free energy:  
\begin{align*}
G_{\beta}(\bm{m})&:=\min_{Q}\extr{\bm{\eta},\gamma}\Big\{\mcal{F}[Q] - \sum_{i \in V}\eta_i\Big[\sum_{\bm{x}}x_i Q(\bm{x}) - m_i\Big]
- \gamma \Big[\sum_{\bm{x}}Q(\bm{x}) - 1\Big] \Big\},
\end{align*}
where the ``extr'' term denotes the extremum with respect to the assigned parameters. 
Through a straightforward manipulation, we obtain $G_{\beta}(\bm{m})$ in the form~\cite{Yasuda2013}  
\begin{align}
G_{\beta}(\bm{m}) 
&=- \beta\bm{h}^{\mrm{T}}\bm{m}
 + \max_{\bm{\eta}}\Big\{ \bm{\eta}^{\mrm{T}}\bm{m} 
 - \ln \sum_{\bm{x}}\exp\big[ \bm{\eta}^{\mrm{T}}\bm{x} +  \beta \bm{x}^{\mrm{T}}\bm{J}\bm{x}/ 2\big]\Big\}.
\label{eqn:GibbsFreeEnergy}
\end{align}
The relation $\beta F:=-\ln Z = \min_{\bm{m}}G_{\beta}(\bm{m})$ holds, where $F$ is the true Helmholtz free energy of the Boltzmann machine. 
The high-temperature expansion of $G_{\beta}(\bm{m})$ is the Plefka expansion~\cite{Plefka1982,Yasuda2012}.

For any $\bm{m}$ and $\beta$, the relation $\ave{x_i}_{\beta}^* = m_i$ holds $\forall i \in V$,  
where $\ave{e(\bm{x})}_{\beta}^*$ is the expectation of $e(\bm{x})$ with respect to 
the distribution $P_{\beta}^*(\bm{x}) \propto \exp[ (\bm{\eta}^*)^{\mrm{T}}\bm{x} +  \beta \bm{x}^{\mrm{T}}\bm{J}\bm{x}/ 2]$.   
This distribution is obtained via the maximum condition for $\bm{\eta}$ in Eq.(\ref{eqn:GibbsFreeEnergy}).
Here, $\bm{\eta}^*$ are the values of the Lagrange multipliers $\bm{\eta}$ in Eq.(\ref{eqn:GibbsFreeEnergy}) 
that satisfy the maximum condition in Eq.(\ref{eqn:GibbsFreeEnergy}).

There is an important relation, referred to as the linear response relation, between the Hessian matrix of the Gibbs free energy, 
$\braket{i|\bm{H}_{\beta}|j} := \partial^2 G_{\beta}(\bm{m})/ \partial m_i \partial m_j$, 
and the susceptibility matrix, $\braket{i|\bm{\chi}_{\beta} |j} := \ave{x_i x_j}_{\beta}^* -m_im_j$, 
where the notation $\braket{i|\bm{A}|j}$ denotes the $(ij)$-th element of the matrix $\bm{A}$. 
The derivative of the equation $m_i = \sum_{\bm{x}}x_i P_{\beta}^*(\bm{x})$ with respect to $m_j$ is 
$\delta_{i,j} = \sum_{k \in V}\braket{i | \bm{\chi}_{\beta} | k}[\partial \eta_k^* / \partial m_j]$, 
where $\delta_{i,j}$ is the Kronecker delta.
On the other hand, the derivative of the Gibbs free energy in Eq.(\ref{eqn:GibbsFreeEnergy}) with respect to $m_i$ is 
$\partial G_{\beta}(\bm{m}) / \partial m_i = \eta_i^*$. 
From the above two equations, we obtain
$\delta_{i,j} 
= \sum_{k \in V}\braket{i | \bm{\chi}_{\beta} | k} \braket{k | \bm{H}_{\beta} | j}$,
and, hence, the linear response relation $\bm{H}_{\beta}^{-1} = \bm{\chi}_{\beta}$ is obtained.

From Taylor's theorem, Eq.(\ref{eqn:GibbsFreeEnergy}) can be expressed as
\begin{align}
G_{\beta}(\bm{m}) =  G_0(\bm{m}) + \int_0^{\beta} \frac{\partial G_t(\bm{m}) }{ \partial t} dt.
\label{eqn:Plefka_GibbsFreeEnergy_TaylorTheorem}
\end{align}
Using the susceptibility matrix, we obtain
\begin{align*}
\frac{\partial G_{\beta}(\bm{m})}{ \partial \beta} = - \bm{h}^{\mrm{T}}\bm{m}  - \ave{\bm{x}^{\mrm{T}}\bm{J}\bm{x}}_{\beta}^* /2
=- \bm{h}^{\mrm{T}}\bm{m} - \bm{m}^{\mrm{T}}\bm{J} \bm{m}/2 -  \trace{(\bm{\chi}_{\beta}\bm{J})} /2.
\end{align*} 
From this equation, Eq.(\ref{eqn:Plefka_GibbsFreeEnergy_TaylorTheorem}), and the linear response relation, we obtain
\begin{align}
G_{\beta}(\bm{m})
= G_0(\bm{m})- \beta\bm{h}^{\mrm{T}}\bm{m} - \frac{\beta}{2} \bm{m}^{\mrm{T}}\bm{J} \bm{m}
-\frac{1}{2}\int_0^{\beta}  \trace{(\bm{H}_{t}^{-1} \bm{J})}dt. 
\label{eqn:Plefka_GibbsFreeEnergy}
\end{align}
The first term in Eq.(\ref{eqn:Plefka_GibbsFreeEnergy}) is
\begin{align}
G_0(\bm{m}) 
&= \sum_{i \in V}\Big[\frac{1+m_i}{2} \ln \frac{1+m_i}{2} +  \frac{1-m_i}{2} \ln \frac{1-m_i}{2}\Big].
\label{eqn:Plefka_GibbsFreeEnergy_beta=0}
\end{align}
Hence, $\braket{i |\bm{H}_0| j} = \delta_{i,j}  [1-m_i^2]^{-1}$. 

In derivation (i) (mentioned in the first paragraph), the Hessian matrix is approximated as 
$\braket{i|\bm{H}_{\beta}|j} \approx \delta_{i,j} V_i(\beta) + \braket{i |\bm{H}_0| j} - \beta J_{ij}$ in Eq.(\ref{eqn:Plefka_GibbsFreeEnergy}), where $V_i(\beta)$ is the variance of the Gaussian-type of cavity field on $i$, 
which is assumed to be independent of $\bm{h}$. 
This approximation is made despite the fact that cavity fields should depend on all of the parameters of the energy functions in general~\cite{Opper&Winther2001a, Opper&Winther2001b}. 
On the other hand, in derivation (ii), as an approximation, the second term in Eq.(\ref{eqn:Plefka_GibbsFreeEnergy_TaylorTheorem}) 
is replaced with $G_{\beta}^{\mrm{Gauss}}(\bm{m}) - G_{0}^{\mrm{Gauss}}(\bm{m})$, 
where $G_{\beta}^{\mrm{Gauss}}(\bm{m})$ is a tractable Gibbs free energy 
originating from a Gaussian distribution with an energy function similar to Eq.(\ref{eqn:energy})~\cite{Opper&Winther2001b, AdaTAP_NIPS14}. 

In the following, we propose a new method for deriving the ATAPFE. 
In the proposed method, we express the fourth term in Eq.(\ref{eqn:Plefka_GibbsFreeEnergy}) by a Gibb free energy of a Gaussian model, 
and the proposed deriving method appears similar to derivation (ii). 
However, the both methods are essentially different. 
In derivation (ii), to obtain the ATAPFE, we approximate $G_{\beta}^{\mrm{Gauss}}(\bm{m})$ by replacing the second term in Eq.(\ref{eqn:Plefka_GibbsFreeEnergy_TaylorTheorem}) with the tractable terms, $G_{\beta}^{\mrm{Gauss}}(\bm{m}) - G_{0}^{\mrm{Gauss}}(\bm{m})$, with no mathematical justification. 
Whereas, in our method, we exactly express $G_{\beta}^{\mrm{Gauss}}(\bm{m})$ in terms of a Gibb free energy of a Gaussian model, 
and we arrive at the ATAPFE as a result of a high-temperature expansion of the exact expression.

\section{Alternative Form of Gibbs Free Energy and Adaptive TAP Free Energy}


For the matrix $\bm{A}_{\beta} := \bm{H}_{\beta} + \beta \bm{J}$, we define the Gaussian type of Helmholtz free energy expressed as
\begin{align}
K_{\beta}(\bm{A}_{\beta}):= - \ln \int_{-\infty}^{\infty}\exp \Big[ -\frac{1}{2} \bm{z}^{\mrm{T}} [\bm{A}_{\beta} - \beta\bm{J}]\bm{z}\Big] d\bm{z},
\label{eqn:FreeEnergyGauss}
\end{align}
where we assume that $\bm{A}_{\beta} - \beta\bm{J} = \bm{H}_{\beta}$ is a positive definite matrix. 
This assumption is convention for the Plefka expansion.
Similar to Eq.(\ref{eqn:Plefka_GibbsFreeEnergy_TaylorTheorem}), we can express the free energy in Eq.(\ref{eqn:FreeEnergyGauss}) as
$
K_{\beta}(\bm{A}_{\beta}) = K_{0}(\bm{A}_{0}) + \int_0^{\beta} [\partial K_t(\bm{A}_{t}) / \partial t] dt
$.
Therefore, we obtain
\begin{align}
&K_{\beta}(\bm{A}_{\beta})= K_{0}(\bm{A}_{0}) -\frac{1}{2}\int_0^{\beta}  \trace{(\bm{H}_{t}^{-1} \bm{J})}dt 
+ \frac{1}{2}\int_0^{\beta} \Ave{ \bm{z}^{\mrm{T}} \frac{\partial \bm{A}_{t} }{\partial t} \bm{z} }_{\bm{H}_{t}}dt,
\label{eqn:Plefka_FreeEnergyGauss}
\end{align}
where $\ave{e(\bm{z})}_{\bm{H}_{\beta}}$ is the expectation of $e(\bm{z})$ with respect to the Gaussian distribution 
$p(\bm{z} \mid \bm{H}_{\beta})\propto \exp[- \bm{z}^{\mrm{T}} \bm{H}_{\beta}\bm{z} / 2]$.

For a test distribution $q(\bm{z})$, we consider the variational free energy, 
\begin{align*}
\mcal{K}[q]:=  \int_{-\infty}^{\infty} 
 [\bm{z}^{\mrm{T}}\bm{H}_{\beta}\bm{z} / 2] q(\bm{z})d\bm{z}
  + \int_{-\infty}^{\infty} q(\bm{z}) \ln q(\bm{z}) d\bm{z}.
\end{align*}
Minimizing the variational free energy with respect to $q(\bm{z})$ under the constraint, 
$\int_{-\infty}^{\infty} z_i^2 q(\bm{z}) d\bm{z} = \xi_i$ $\forall i \in V$, we obtain the Gibbs free energy as 
\begin{align*}
\Phi_{\beta}(\bm{\xi})&:=\min_{q}\extr{\bm{\lambda},\gamma} \Big\{\mcal{K}[q] 
+ \frac{\beta}{2}\sum_{i \in V}\lambda_i \Big[\int_{-\infty}^{\infty} z_i^2 q(\bm{z}) d\bm{z} - \xi_i\Big] - \gamma \Big[\int_{-\infty}^{\infty}  q(\bm{z}) d\bm{z} - 1\Big] \Big\}.
\end{align*}
This can be reduced to
\begin{align*}
\Phi_{\beta}(\bm{\xi})
=\frac{1}{2}\max_{\bm{\lambda}}\big\{-\beta\bm{\lambda}^{\mrm{T}}\bm{\xi}+ \ln \det [\bm{H}_{\beta} + \beta\bm{\Lambda}]\big\}
-c,
\end{align*}
where $\bm{\Lambda}$ is a diagonal matrix with diagonal elements $\{\lambda_i\}$ and $c := [|V| / 2] \ln [2\pi]$. 
Because $K_{\beta}(\bm{A}_{\beta}) = \min_{\bm{\xi}}\Phi_{\beta}(\bm{\xi})$ and, at the minimum point, $\xi_i = \braket{i|\bm{H}_{\beta}^{-1}|i} = 1 - m_i^2$ $\forall i \in V$, we obtain
\begin{align}
K_{\beta}(\bm{A}_{\beta}) 
& = \frac{1}{2} \max_{\bm{\lambda}}\Big\{-\beta\sum_{i \in V}\lambda_i [1-m_i^2]
+ \ln \det [\bm{H}_{\beta} + \beta\bm{\Lambda}]\Big\} -c.
\label{eqn:FreeEnergyGauss_Transformed}
\end{align}
When $\beta = 0$, Eq.(\ref{eqn:FreeEnergyGauss_Transformed}) is reduced to
\begin{align}
K_{0}(\bm{A}_{0}) = -\frac{1}{2}\sum_{i \in V} \ln [1-m_i^2]-c.
\label{eqn:FreeEnergyGauss_Transformed_beta=0}
\end{align}

From Eqs.(\ref{eqn:Plefka_GibbsFreeEnergy}), (\ref{eqn:Plefka_GibbsFreeEnergy_beta=0}), 
(\ref{eqn:Plefka_FreeEnergyGauss})--(\ref{eqn:FreeEnergyGauss_Transformed_beta=0}), 
we obtain the alternative form of the Gibbs free energy as
\begin{align}
G_{\beta}(\bm{m})&=\sum_{i \in V}\Big[\frac{1+m_i}{2} \ln \frac{1+m_i}{2} +  \frac{1-m_i}{2} \ln \frac{1-m_i}{2}\Big]
- \beta\bm{h}^{\mrm{T}}\bm{m} - \frac{\beta}{2} \bm{m}^{\mrm{T}}\bm{J} \bm{m} \nn
&+\frac{1}{2} \max_{\bm{\lambda}}\Big\{-\beta\sum_{i \in V}\lambda_i [1-m_i^2]
+ \ln \det [\bm{H}_{\beta} + \beta\bm{\Lambda}]\Big\} +\frac{1}{2}\sum_{i \in V} \ln [1-m_i^2]
- \frac{R_{\beta}}{2},
\label{eqn:AlternativeGibbsFreeEnergy}
\end{align}
where 
$R_{\beta}:=\int_0^{\beta} \ave{ \bm{z}^{\mrm{T}} [\partial \bm{A}_{t}  / \partial t] \bm{z} }_{\bm{H}_{t}} dt$.
The expression in Eq.(\ref{eqn:AlternativeGibbsFreeEnergy}) is exact when $\bm{H}_{\beta}$ is positive definite.

In the following, we obtain the ATAPFE via a high-temperature approximation of the intractable Hessian matrix $\bm{H}_{\beta}$ in the $G_{\beta}(\bm{m})$ expression given in Eq.(\ref{eqn:AlternativeGibbsFreeEnergy}). 
Using the Plefka expansion~\cite{Plefka1982}, 
\begin{align*}
G_{\beta}(\bm{m})= G_0(\bm{m})- \beta\bm{h}^{\mrm{T}}\bm{m} - \beta  \bm{m}^{\mrm{T}}\bm{J} \bm{m} / 2 + O(\beta^2),
\end{align*}
$\bm{H}_{\beta}$ can be expanded as $\bm{H}_{\beta} = \bm{H}_0 - \beta \bm{J} + O(\beta^2)$. 
Thus, $\bm{H}_{\beta} \approx \bm{H}_0 - \beta \bm{J}$ when $\beta\ll 1$. 
This high-temperature approximation corresponds to the na{\"i}ve mean-field approximation of $\bm{H}_{\beta}$.
Further, for this high-temperature approximation, the remainder $R_{\beta}$ vanishes, because 
$\bm{A}_{\beta} =  \bm{H}_{\beta} + \beta \bm{J} = \bm{H}_{0} + O(\beta^2)$ and therefore $\partial \bm{A}_{\beta} / \partial \beta \approx  \bm{0}$. 
From these approximations, we arrive at  
\begin{align}
G_{\beta}(\bm{m})&\approx \sum_{i \in V}\Big[\frac{1+m_i}{2} \ln \frac{1+m_i}{2} +  \frac{1-m_i}{2} \ln \frac{1-m_i}{2}\Big]
- \beta\bm{h}^{\mrm{T}}\bm{m} - \frac{\beta}{2} \bm{m}^{\mrm{T}}\bm{J} \bm{m}\nn
&+\frac{1}{2} \max_{\bm{\lambda}}\Big\{-\beta\sum_{i \in V}\lambda_i [1-m_i^2]
+ \ln \det \beta[\bm{\Lambda} - \bm{J}]\Big\} +\frac{1}{2}\sum_{i \in V}\Big\{1 +  \ln [1-m_i^2]\Big\},
\label{eqn:AdaptiveTAPFreeEnergy}
\end{align}
where we redefine $\bm{\Lambda} \leftarrow \bm{\Lambda} + \beta^{-1}\bm{H}_0$. 
The expression in Eq. (\ref{eqn:AdaptiveTAPFreeEnergy}) coincides with the ATAPFE presented in Ref.\cite{Opper&Winther2001b}.
From the proposed derivation, we can understand that the ATAPFE is justified when the Hessian matrix $\bm{H}_{\beta}$ can be expressed by $\bm{H}_0 - \beta \bm{J}$.

\section{Conclusion}

In this paper, we have provided a new method for deriving the ATAPFE. 
The proposed derivation allows the ATAPFE to be obtained via a high-temperature expansion of the Hessian matrix in the Gibbs free energy with no special assumptions, 
and facilitates to obtain higher-order approximations. 
For example, the Hessian matrix can be expanded as~\cite{Yasuda2007}
\begin{align*} 
\braket{i|\bm{H}_{\beta}|j} = \delta_{i,j} [1-m_i^2]^{-1}- \beta J_{ij} - 2\beta^2 J_{ij}^2 m_im_j
+\beta^2 \delta_{i,j} \sum_{k \in V}J_{ik}^2[1-m_k^2] + O(\beta^3), 
\end{align*}
and this expression will yields a high-order approximation of the ATAPFE. 

Note that this method is not directly applicable to multivalued cases, because the $x_i^2 = 1$ relation is crucial in such scenarios. 
This is a limitation of our current method,  
and extension of this technique to multivalued cases is a topic for future research.

\subsubsection*{acknowledgment}
This work was partially supported by CREST, Japan Science and
Technology Agency and by JSPS KAKENHI (Grant Numbers 15K00330, 25280089, and 15H03699).

\bibliographystyle{jpsj}
\bibliography{cite}

\begin{thebibliography}{1}

\bibitem{Opper&Winther2001a}
M.~Opper and O.~Winther: Phys. Rev. Lett. {\bfseries 86} (2001) 3695.

\bibitem{Opper&Winther2001b}
M.~Opper and O.~Winther: Phys. Rev. E {\bfseries 64} (2001) 056131.

\bibitem{Yasuda2013}
M.~Yasuda and K.~Tanaka: Phys. Rev. E {\bfseries 87} (2013) 012134.

\bibitem{AdaTAP_NIPS14}
L.~Csat{\'{o}}, M.~Opper, and O.~Winther: In Advances in Neural Information
  Processing Systems 14  (2001) 657.

\bibitem{Plefka1982}
T.~Plefka: J. Phys. A: Math. and Gen. {\bfseries 15} (1982) 1971.

\bibitem{Yasuda2012}
M.~Yasuda and K.~Tanaka: Philosophical Magazine {\bfseries 92} (2012) 192.

\bibitem{Yasuda2007}
M.~Yasuda and K.~Tanaka: J. Phys. A: Math. and Theor. {\bfseries 40} (2007)
  9993.

\end{thebibliography}

\end{document}